\documentclass[sigconf,10pt]{acmart}

\renewcommand\footnotetextcopyrightpermission[1]{} 
\usepackage{diagbox}
\usepackage[english]{babel}
\usepackage{color,colortbl,multirow,hyphenat}
\usepackage{tikz,hyperref,xspace,booktabs}
\usepackage{tabularx,enumitem}
\usepackage[labelformat=simple, textfont=normalfont]{subcaption}
\usepackage{multirow}
\usepackage[normalem]{ulem}
\usepackage{booktabs,amsfonts,algorithm}
\usepackage{titlesec}

\usepackage{tikz}
\usetikzlibrary{quantikz}
\usepackage{mathtools}
\usepackage[math]{cellspace}
\usepackage{physics}
\usepackage{blindtext}
\usepackage[T1]{fontenc}
\usepackage[utf8]{inputenc}
\usepackage{newtxmath,bm,courier}
\usepackage{xspace,enumitem,fancyhdr,fancyref,wrapfig}
\usepackage{amsmath,amsfonts,bm}
\usepackage{tabularx,tablefootnote,wrapfig,hyphenat}
\usepackage[labelformat=simple,skip=0pt]{subcaption}
\usepackage{xfrac,sparklines}
\usepackage{algorithm}
\usepackage{algpseudocode}

\usepackage{tabularx}
\usepackage{wrapfig}
\usepackage{lipsum}
\usepackage{mathtools}
\usepackage{amsmath}
\usepackage{upgreek}
\usepackage{float}
\usepackage{hyperref,balance}
\usepackage{listings} 
\usepackage{natbib}

\newcommand{\tens}[1]{%
  \mathbin{\mathop{\otimes}\displaylimits_{#1}}%
}

\usepackage[toc]{appendix}

\newcommand{\shortname}{QGateD-Polar}
\newcommand{\systemname}{\shortname{}}
\newcommand{\shortnames}{QGateD-Polar's}
\newcommand{\systemnames}{\shortnames{}}

\setitemize{itemsep=3pt,topsep=4pt,parsep=2pt,partopsep=0pt,leftmargin=1.5em}
\setenumerate{itemsep=3pt,topsep=4pt,parsep=1pt,partopsep=0pt,leftmargin=1.5em}
\definecolor{Gray}{gray}{0.9}
\definecolor{LightGreen}{rgb}{0.88,1,0.88}
\definecolor{LightYellow}{rgb}{1,1,0.8}
\definecolor{LightOrange}{rgb}{1,0.85,0.8}
\definecolor{LightRed}{rgb}{1,0.80,0.80}

\newcolumntype{P}[1]{>{\centering\arraybackslash}p{#1}}
\newcolumntype{M}[1]{>{\centering\arraybackslash}m{#1}}

\acmDOI{}

\newcommand{\parahead}[1]{\vspace*{1ex minus 0.25ex}\noindent %
  {\bfseries #1.}}
\setitemize{itemsep=3pt,topsep=4pt,parsep=2pt,partopsep=0pt,leftmargin=1.5em}
\setenumerate{itemsep=3pt,topsep=4pt,parsep=1pt,partopsep=0pt,leftmargin=1.5em}

\begin{document}

\fancyfoot{}
\setcopyright{none}
\renewcommand\footnotetextcopyrightpermission[1]{}
\pagestyle{plain}
\thispagestyle{empty}

\title{\vspace{-20pt} Decoding Polar Codes via Noisy Quantum Gates: Quantum Circuits and Insights\\[-1.0ex]}
\affiliation{\Large Srikar Kasi$^{\star, \dagger}$, John Kaewell$^{\dagger}$, Shahab Hamidi-Rad$^{\dagger}$, Kyle Jamieson$^{\star}$}
\affiliation{\large $^\star$Princeton University, $^{\dagger}$InterDigital, Inc.\\[2.0ex]}

\begin{abstract}

The use of quantum computation for wireless network applications is emerging as a promising paradigm to bridge the performance gap between in-practice and optimal wireless algorithms. While today's quantum technology offers limited number of qubits and low fidelity gates, application-based quantum solutions help us to understand and improve the performance of such technology even further. This paper introduces \textbf{\shortname{}}, a novel Quantum Gate-based Maximum-Likelihood Decoder design for Polar error correction codes, which are becoming widespread in today's 5G and tomorrow's NextG wireless networks. QGateD-Polar uses quantum gates to dictate the time evolution of Polar code decoding---from the received wireless soft data to the final decoded solution---by leveraging quantum phenomena such as superposition, entanglement, and interference, making it amenable to quantum gate-based computers. Our early results show that QGateD-Polar achieves the Maximum Likelihood performance in ideal quantum simulations, demonstrating how performance varies with noise.

\end{abstract}

\maketitle

\section{Introduction}
\label{s: intro}

Today's 4G and 5G cellular wireless networks are experiencing unprecedented growth in traffic at base stations due to increased subscriber numbers and their higher quality of service requirements \cite{cisco}. Consequently, associated baseband units (BBUs) must handle data at high computational throughputs to enable high-spectral-efficiency networks envisioned in 5G and NextG communication systems \cite{kasi2021challenge}.

A key component of a cellular BBU processing is the Error Correction Code (ECC), which seeks to correct the bit errors that the wireless channel inevitably introduces into the data transmission. While several high performance ECCs that achieve Shannon capacity exist today, realizing their capacity-achieving benefits in practice has been challenging due to their computationally-heavy decoder processing requirements \cite{kasi2020towards, hailes2015survey}. In particular, traditional ECC decoders backed by complementary metal oxide semiconductor (CMOS)-based computation entail two major problems:
\textbf{(1)} Since all the received wireless data must pass through an ECC decoder, it must operate at high clock frequency to achieve high computational throughput. However, the conventional tradeoff between decoder accuracy and throughput significantly constrains the decoder operational clock frequency \cite{hailes2015survey}, while on the other hand CMOS hardware advancements are not maintaining the pace they had in the past years due to transistor sizes approaching atomic limits and the end of Moore's Law (expected \textit{ca.} 2025-2030). \textbf{(2)} The power consumption of CMOS-based ECC decoders largely depends on the amount of computation at hand, so it is expected to significantly increase with the increasing user demand in cellular wireless networks. These issues therefore call into question the prospects of CMOS-based ECC decoders to handle NextG cellular networks' user demand without imposing any spectral and energy efficiency bottlenecks \cite{kasi2021challenge}.

To address these issues, researchers are beginning to investigate alternate approaches to CMOS-based ECC decoders. Recent prior work in this area has explored the Quantum Annealing (QA) approach, a specialized quantum technology tailored for optimization problems, for decoding several ECCs such as Low Density Parity Check (LDPC) codes, Hamming codes, among others \cite{kasi2020towards, ide2020maximum, bian2014discrete}. While these approaches have shown promising results, they demand significantly huge amount of computational resources (\textit{e.g.,} qubits) which are expensive and unlikely to be available in the near-term \textit{noisy intermediate scale quantum} (NISQ) era. For instance, a QA-based LDPC decoder requires about 21K logical qubits for decoding a single 5G-NR LDPC code \cite{kasi2020towards}. Further issues surrounding qubit quality, coherence times, and readout fidelity increase these requirements even further.

In order to overcome these limitations, this paper investigates Quantum Gate technology for ECC decoding, which is widely recognized as the universal method for quantum computing \cite{sleator1995realizable}. While there exist many interesting ECC candidates to study, in this paper we focus on \textit{Polar~Codes}, the first class of error correction codes that provably achieve the symmetric capacity of any binary-input discrete memoryless channel \cite{arikan2009channel}. Polar Codes stand out today because of their exceptional error correcting capability, low error floor, simple encoding, among others. However, despite these desirable properties, Polar Codes face several practical challenges if they are to manage their decoder design complexity while at the same time maintaining their capacity-achieving properties. The state-of-the-art \textit{Successive Cancellation List} (SCL) decoder can achieve capacity asymptotically but at the price of high complexity and high latency. Furthermore, the performance of finite search SCL decoders is unsatisfactory for code block lengths used in practice. Due to these issues, the Polar Code use in 5G New Radio is currently limited to control channels with short block lengths \cite{3gpp}. But Polar Codes' solid theoretical foundation, simple encoder implementation, and adjustable code rate would make them viable candidates for high-speed data channels, if the decoder latency could be minimized while simultaneously maintaining a low bit error rate and near-capacity-limit rate performance.

It is with this vision in mind that this paper introduces \textbf{\systemname{}}, the first \textbf{Q}uantum \textbf{Gate}-based Maximum-Likelihood \textbf{D}ecoder design for \textbf{Polar} error correction codes. \systemname{} sets aside traditional boolean logic gates-based computational methods, instead proposes a fresh quantum gates-based computational framework for decoding Polar codes. This would open up new possibilities in the design of Polar code decoders, thus overcoming the aforementioned hurdles to adoption. \systemname{} works by initializing a system of qubits in a quantum superposition state characterized by the received wireless soft data (with channel noise), then guides the evolution of this quantum state over time via a series of quantum entanglement operations such that these operations mimic the computational behavior of the Polar Code's Generator matrix. \systemname{} next amplifies the quantum states that agree with the Polar encoding conditions while simultaneously suppressing the ones that do not agree, via quantum interference. The quantum bits are then measured into classical registers (see \S\ref{s: design}).

We have evaluated \systemname{} for eight bit, half rate Polar codes in both ideal and noisy quantum gate simulations. Our results show that \systemname{} achieves the Maximum-Likelihood performance in the ideal scenario, whereas about half an order of bit error rate magnitude degrades under \textit{BitFlip} quantum noise with an error probability of $10^{-3}$. Our noisy simulation experiments are preliminary and are evaluated under sub-optimal parameters, and so there is more room for performance improvement. We provide reasoning for this performance degradation and discuss challenges involved therein, towards the goal of our eventual full system design for decoding long 1,024 bit 5G-NR Polar codes.

\section{Primer: Quantum Gate Computation}
\label{s: primer-quantum}

This section provides necessary background and fundamentals of quantum gate computation.

\subsection{Quantum Bits and States}
\label{s: qbits}

\begin{figure}
\centering
\includegraphics[width=0.4\linewidth]{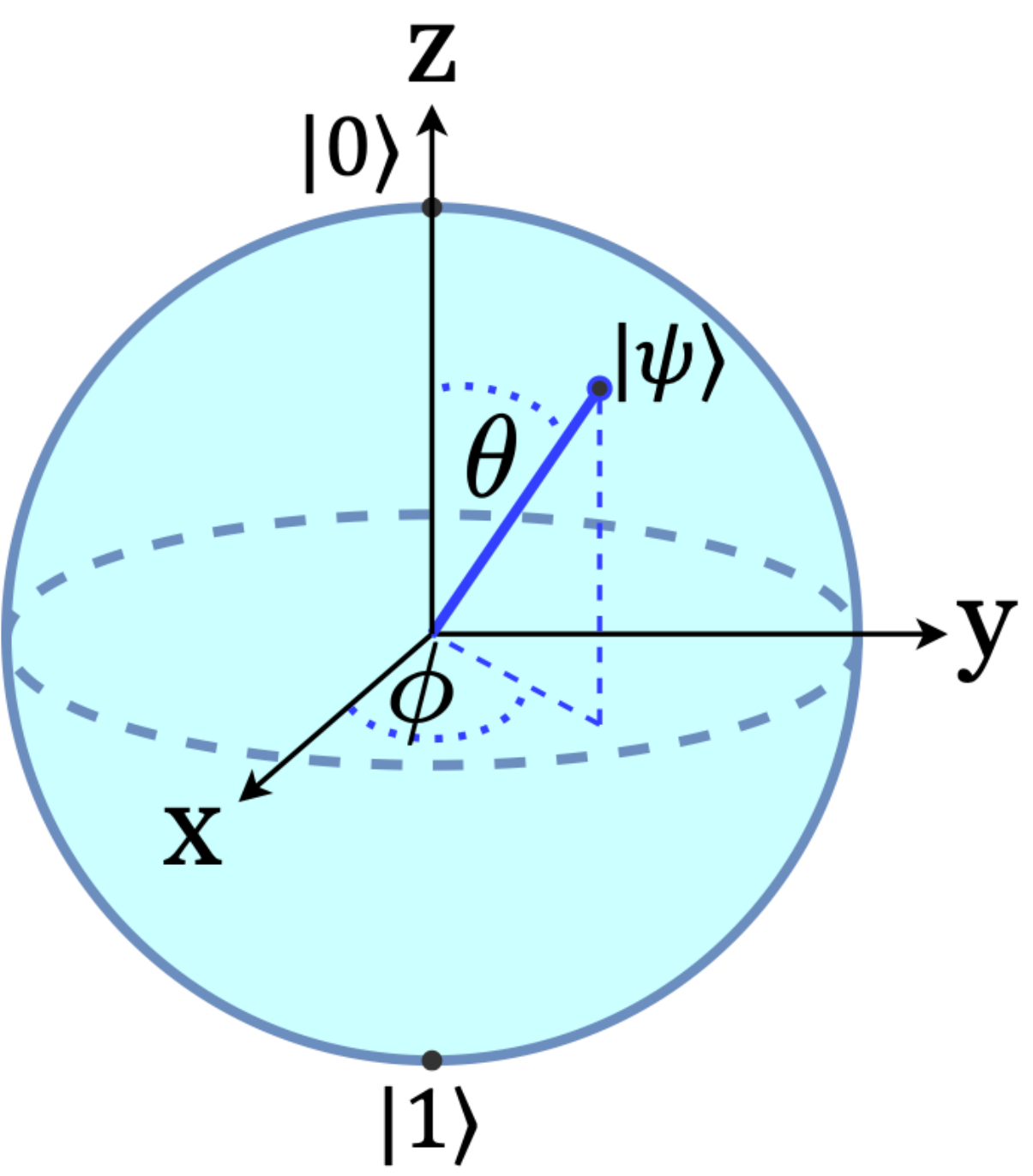}
\caption{Bloch Sphere representation of a qubit (\S\ref{s: qbits}).}
\label{fig: qubit}
\end{figure}

While classical computation uses bits to represent information, quantum computation uses qubits, physical devices that can exist in a quantum \textit{superposition} state. The state $\ket{\psi}$ of a single qubit is mathematically expressed as:
\begin{equation}
    \ket{\psi} = \alpha\ket{0} + \beta\ket{1} = \begin{pmatrix}
        \alpha\\ \beta
    \end{pmatrix}
    \label{eqn: state}
\end{equation}
where $\alpha$ and $\beta$ are complex-valued numbers, called probability amplitudes, that satisfy the condition $|\alpha|^2 + |\beta|^2 = 1$, and $\ket{0} = \begin{pmatrix}
    1\\0
\end{pmatrix}$ and $\ket{1} = \begin{pmatrix}
    0\\1
\end{pmatrix}$ are the short-hand Dirac notations of the qubit's \textit{computational basis}. Measuring a qubit in state $\ket{\psi}$ yields a value of zero ($\ket{0}$) with probability $|\alpha|^2$ and one ($\ket{1}$) with probability $|\beta|^2$. To visualize the state of a qubit, we can convert its description in Eq.~\ref{eqn: state} into the form:
\begin{equation}
    \ket{\psi} = cos(\theta/2)\ket{0} + e^{i\phi}sin(\theta/2)\ket{1}
\end{equation}
where $\theta$ and $\phi$ are some real-valued parameters. By interpreting these parameters as angles of a spherical co-ordinate system with radial distance of one, we can plot the state of a single qubit on the surface of a unit sphere know as a \textit{Bloch Sphere}. Figure~\ref{fig: qubit} illustrates the Bloch sphere representation of a qubit with $\ket{0}$ and $\ket{1}$ as its north and south poles, respectively. The angles $\theta$ and $\phi$ capture the \textit{relative amplitude} and \textit{relative phase} between the $\ket{0}$ and $\ket{1}$ basis states.

\begin{figure*}[ht]
\centering
\begin{subfigure}[b]{0.32\linewidth}
\includegraphics[width=\linewidth]{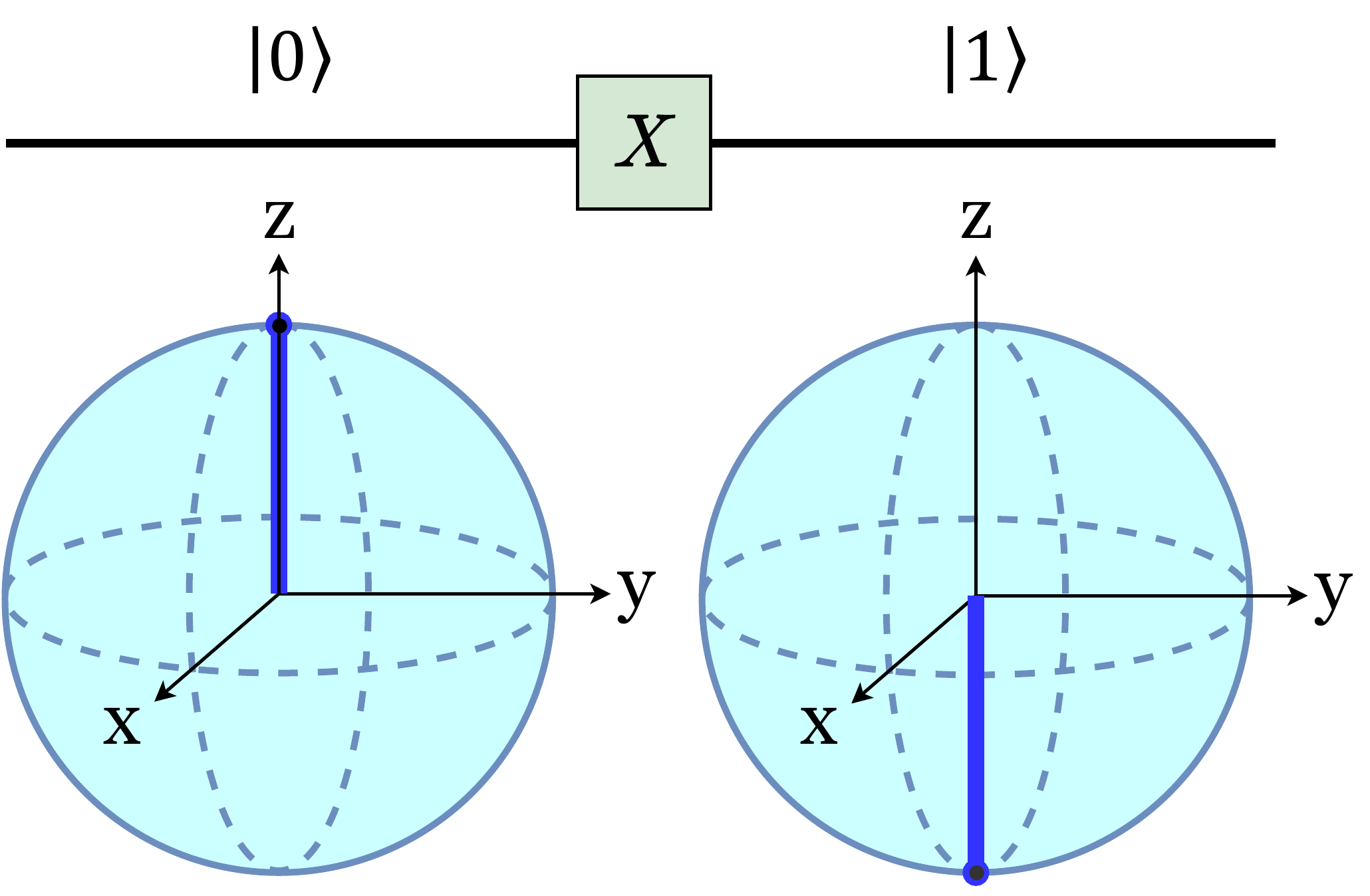}
\caption{X Gate.}
\label{fig: X}
\end{subfigure}
\hfill
\begin{subfigure}[b]{0.32\linewidth}
\includegraphics[width=\linewidth]{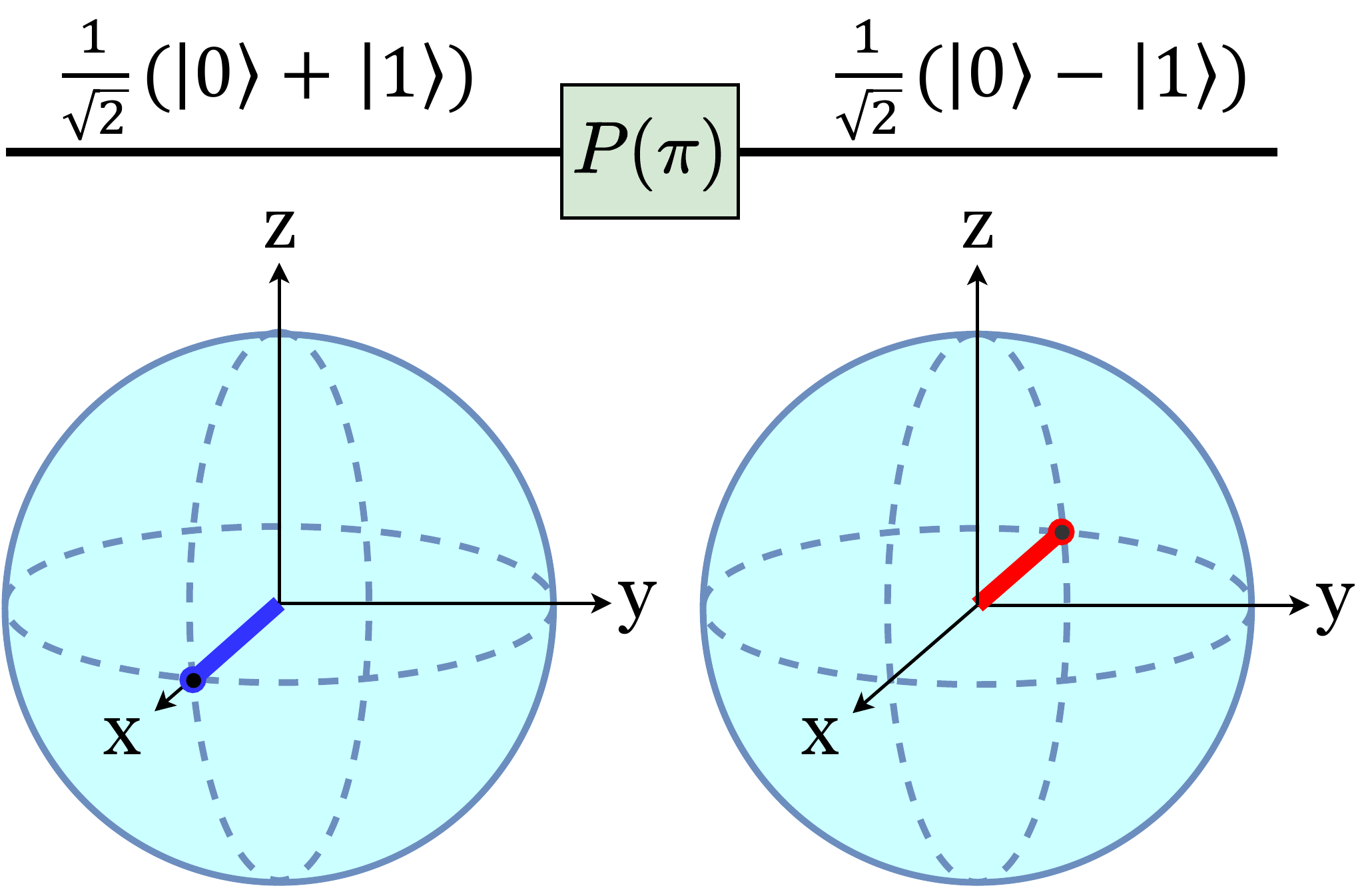}
\caption{P gate.}
\label{fig: P}
\end{subfigure}
\hfill
\begin{subfigure}[b]{0.32\linewidth}
\centering
\includegraphics[width=\linewidth]{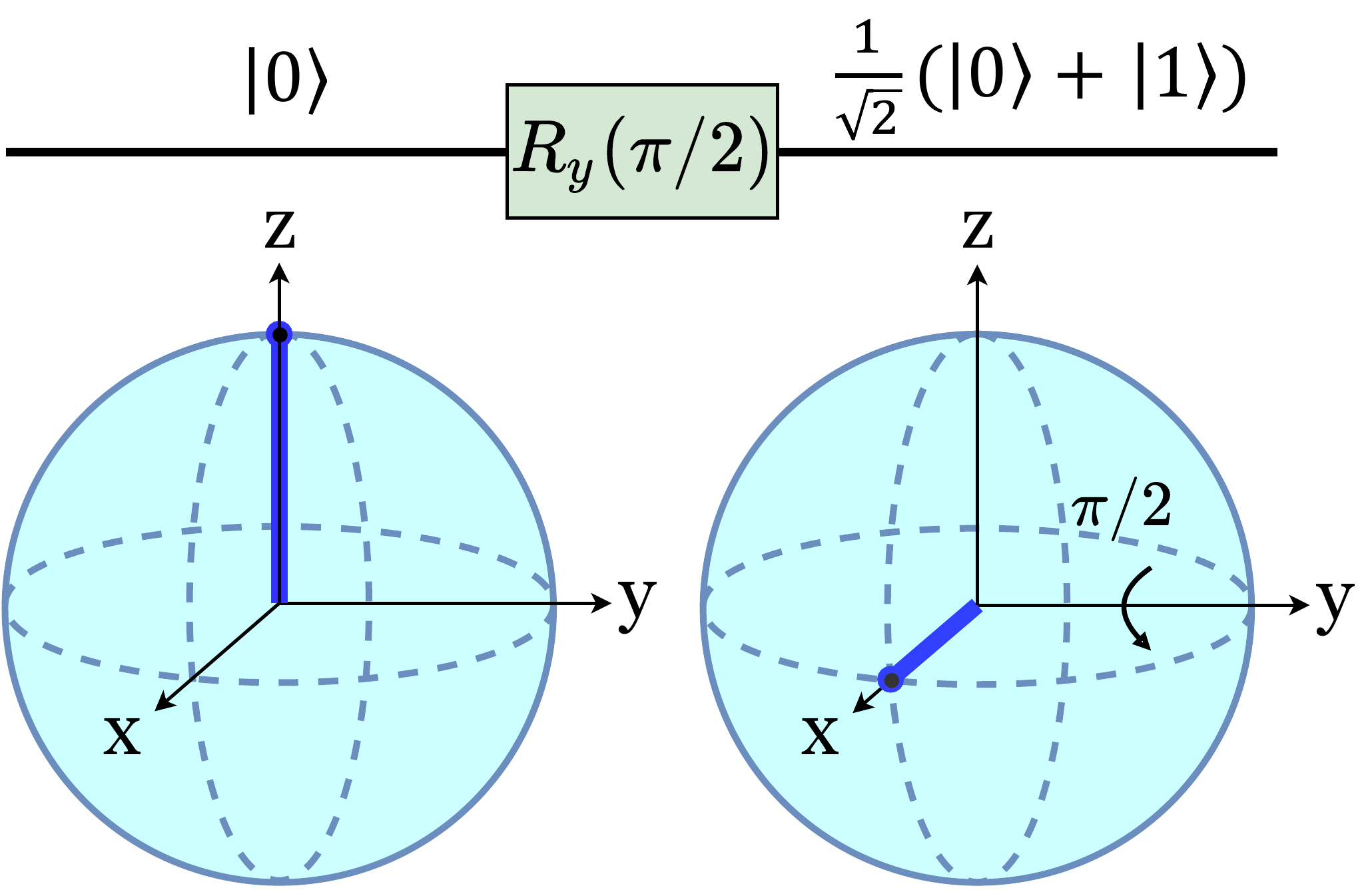}
\caption{$R_y$ gate.}
\label{fig: R}
\end{subfigure}
\caption{An example illustration showing how the state of a single qubit changes under the action of X, P($\lambda$), and RY($\theta$) gates. Blue and red lines indicate a relative phase of zero and $\pi$ radians between the qubit's basis states ($\ket{0}$ and $\ket{1}$).}
\label{fig: example}
\end{figure*}

\parahead{System of qubits} The quantum state of a system of qubits resides in the tensor product of their individual qubit states. Say we have two qubits whose states are $\ket{\psi_0} = \alpha_0\ket{0} + \beta_0\ket{1}$ and $\ket{\psi_1} = \alpha_1\ket{0} + \beta_1\ket{1}$, then the collective state of this two-qubit system is given by $\ket{\psi} = \ket{\psi_0}\otimes \ket{\psi_1}$:
\begin{equation}
    \ket{\psi} = \alpha_0\alpha_1\ket{00} + \alpha_0\beta_1\ket{01} + \beta_0\alpha_1\ket{10} + \beta_0\beta_1\ket{11}
\end{equation}
Measuring this system yields the qubit configuration $\ket{00}$ with probability $|\alpha_0\alpha_1|^2$, and so forth. A similar mathematical formalism applies to higher dimensional qubit systems.

\parahead{Entangled states} It is also possible for a quantum system to exist in a state which cannot be expressed as a tensor product of individual qubit states. Such quantum states are called \textit{entangled} states. An example of a two\nobreakdash-qubit quantum entangled state is the well-known Bell state:
\begin{equation}
    \ket{\psi} = \frac{1}{\sqrt{2}}\Big(\ket{00} + \ket{11}\Big)
    \label{eqn: entangled}
\end{equation}
Note that for any values of $\alpha_0, \alpha_1, \beta_0, \beta_1$, the quantum state in Eq.~\ref{eqn: entangled} cannot be expressed as $(\alpha_0\ket{0} + \beta_0\ket{1})\otimes (\alpha_1\ket{0} +~\beta_1\ket{1})$. The existence of entangled states in the real-world is a physical fact that has important consequences in quantum information. Indeed, without the existence of such states quantum computing would have limited applicability \cite{abhijith2018quantum}.

\subsection{Quantum Gates}
\label{s: gates}

Quantum Gates can be used to manipulate quantum bits as in boolean logic for binary bits. They are \textit{unitary operators} which can act upon one or more qubits, and whose action results in a \textit{unitary transformation} of a quantum state. In a nutshell, a quantum gate described by a matrix $U$ satisfies the unitary matrix property: $UU^H = U^HU = I$, where $U^H$ is the conjugate transpose of $U$ and $I$ is the identity matrix, and a state $\ket{\psi}$ evolves under the action of $U$ as: $\ket{\psi}\longrightarrow U\ket{\psi}$. We here describe a few gates that are necessary to understand our \systemname{} design (described later in \S\ref{s: design}).

\parahead{Single-qubit gates} A single-qubit gate can be used to manipulate superposition of a single qubit. We here describe X, P, and $R_y$ gates whose unitary matrices are:
\vspace{-5pt}
\begin{equation}
    X = \begin{pmatrix}
        0 & 1\\1& 0
    \end{pmatrix}, \hspace{1pt} P (\lambda) = \begin{pmatrix}
        1 & 0\\ 0& e^{i\lambda}
    \end{pmatrix}, \hspace{1pt} R_y(\theta) = \begin{pmatrix}
        cos \frac{\theta}{2} & -sin \frac{\theta}{2}\\ sin \frac{\theta}{2} & cos \frac{\theta}{2}
    \end{pmatrix}
    \label{eqn: matrices}
\end{equation}

\vspace{-5pt}
\textit{X} gate is the quantum analogy of classical NOT gate. It transforms the basis states $\ket{0}$ and $\ket{1}$ to $\ket{1}$ and $\ket{0}$ respectively, and so a state $\ket{\psi} = \alpha\ket{0} + \beta\ket{1}$ evolves under the action of X gate as: $\ket{\psi} \longrightarrow X\ket{\psi} = \beta\ket{0} + \alpha\ket{1}$. It follows:
\begin{equation}
    \ket{\psi}\longrightarrow X\ket{\psi} = \begin{pmatrix}
        0 & 1\\1& 0
    \end{pmatrix}\begin{pmatrix}
        \alpha\\\beta
    \end{pmatrix} = \begin{pmatrix}
        \beta\\\alpha
    \end{pmatrix} = \beta\ket{0} + \alpha\ket{1}
\end{equation}

\textit{P} gate works by creating a relative phase of $\lambda$ radians between $\ket{0}$ and $\ket{1}$ basis states, where $\lambda$ is a parameter. It transforms $\ket{1}$ to $e^{i\lambda}\ket{1}$, and so $\ket{\psi}$ evolves under the action of P gate as: $\ket{\psi}\longrightarrow P(\lambda)\ket{\psi} = \alpha\ket{0} + e^{i\lambda}\beta\ket{1}$. If $\lambda = \pi$, then P gate is equivalent to a more familiar \textit{Z gate}.

$R_y$ gate rotates the state $\ket{\psi}$ by an angle $\theta$ in clockwise direction about the \textit{y-axis} of Bloch sphere. It transforms: $\ket{0}$ to $ cos \frac{\theta}{2} \ket{0} + sin \frac{\theta}{2} \ket{1}$  and $\ket{1}$ to $-sin \frac{\theta}{2} \ket{0} + cos \frac{\theta}{2} \ket{1}$. Figure~\ref{fig: example} illustrates an example, showing how the state of a single qubit changes under X, P($\lambda$), and $R_y$($\theta$) gate operations.

\parahead{Multi-qubit gates} While single-qubit gates manipulate superposition, multi-qubit gates manipulate entanglement among two or more qubits. We here describe Controlled\hyp{}NOT (CNOT) and Multi Controlled-Phase (MCPhase) gates. CNOT gate operates on a \textit{control} and \textit{target} qubit pair. When the control qubit is in state $\ket{1}$, it performs X gate operation on the target qubit. MCPhase gate operates on multiple control qubits and a target qubit. When all the control qubits are in state $\ket{1}$, it performs P($\lambda$) gate operation on the target qubit. To understand this, let us consider an example state:
\begin{equation}
    \ket{\psi} = \frac{1}{\sqrt{2}} \Big( \ket{010} + \ket{111} \Big)
    \label{eqn: cnot_0}
\end{equation}
Operating a CNOT gate on $\ket{\psi}$ with second qubit as control and third qubit as target would result in the transformation:
\begin{equation}
    \ket{\psi}\longrightarrow CNOT_{2,3}\ket{\psi} = \frac{1}{\sqrt{2}} \Big( \ket{011} + \ket{110} \Big)
    \label{eqn: cnot}
\end{equation}
This is because in $\ket{\psi}$ the second qubit (control) is in state $\ket{1}$ in both the basis states $\ket{010}$ and $\ket{111}$, and so the third qubit (target) is flipped in both cases (see Eq.~\ref{eqn: cnot}).
Along a similar argument, we can observe that operating an MCPhase($\lambda$) gate on $\ket{\psi}$ with first and second qubits as controls and third qubit as target would result in the transformation:
\begin{equation}
    \ket{\psi}\longrightarrow MCPhase_{\{1,2\},3}\ket{\psi} = \frac{1}{\sqrt{2}} \Big( \ket{010} + e^{i\lambda}\ket{111} \Big)
    \label{eqn: mcphase}
\end{equation}
Essentially, an MCPhase($\lambda$) gate operating on $n$ qubits transforms the state $\ket{1}^{\otimes n}$ to $e^{i\lambda}\ket{1}^{\otimes n}$ irrespective of the permutation in which the control and target qubits are in. For example, note that MCPhase($\lambda$) gate operating on $\ket{\psi}$ with second and third qubits as controls and first qubit as target would result in the same transformation shown in Eq.~\ref{eqn: mcphase}.

\section{Design}
\label{s: design}

We now introduce our \systemname{} decoder design, beginning with a short description of Polar code encoding.

\begin{figure}
\centering
\includegraphics[width=0.8\linewidth]{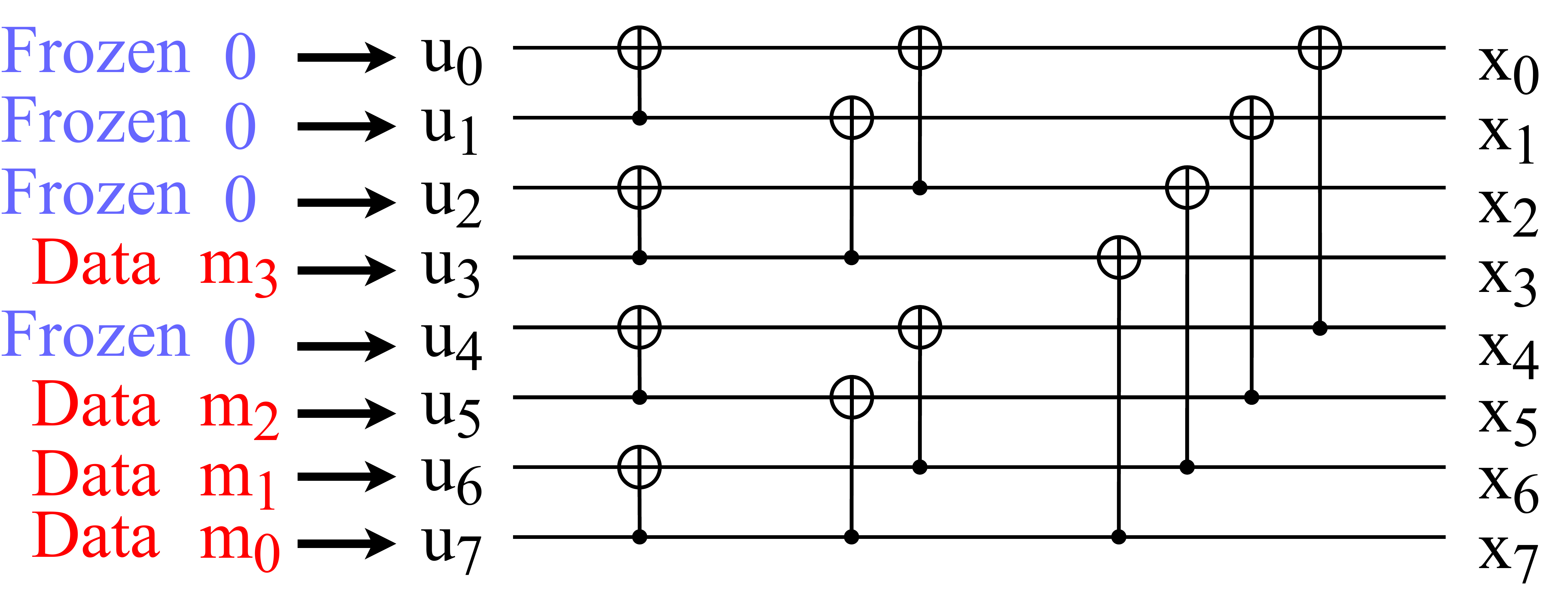}
\caption{Encoding process of an example 8-bit, 0.5 rate Polar code. $m_i$ is a message bit, $u_i$ is an encoder input bit, and $x_i$ is an encoded bit that is transmitted over a wireless channel. $\oplus$ represents classical XOR operation. \vspace{-10pt}}
\label{fig: encoder}
\end{figure} 

A Polar code is functionally described by a generator matrix $\textbf{G}_N = \textbf{G}_2^{\otimes d}$, where $\textbf{G}_2 = \big[\begin{smallmatrix}
  1 & 0\\
  1 & 1
\end{smallmatrix}\big]$ and $\otimes d$ is \textit{d}\hyp{}fold Kronecker product. Let \textit{N} be the code block length adapted for transmitting a message \textbf{m} = $[m_0, m_1, ..., m_{K-1}]$ of length $K \leq N$ bits. Construct the encoder \textit{input vector} \textbf{u} = $[u_0, u_1, ..., u_{N-1}]$ by assigning bits in \textbf{m} to \textit{K} most reliable locations in \textbf{u}, and set the remaining $N-K$ bits to a zero-value. The $u_i$'s that are assigned to zero-value are called \textit{frozen bits}. The encoded codeword is then \textbf{x} =~\textbf{u}$\textbf{G}_N$, which is next interleaved and transmitted over a wireless channel. Figure~\ref{fig: encoder} shows the encoding process of an example 8-bit, 0.5 rate Polar code.

Let \textbf{y} = $[y_0, y_1, ..., y_{N-1}]$ be the respective received wireless soft data corresponding to bits in \textbf{x}, and let \textbf{q} = $[q_0, q_1, ..., q_{N-1}]$ be the qubits \systemname{} uses to decode bits in \textbf{u} respectively. User message \textbf{m} is a portion of \textbf{u} (see Fig.~\ref{fig: encoder}).

\subsection{\systemname{}: Algorithm}
\label{s: algorithm}

Our \systemnames{} decoding algorithm has three steps: \textit{Initialization}, \textit{Reverse Traversal}, and \textit{Frozen Bit Satisfaction}, which we describe here.

\textbf{Initialization.} For decoding a code of block length $N$ bits, we begin with $N$ qubits in the state $\ket{0}^{\otimes N}$. Each qubit $q_i$ is next initialized using an $R_y(\theta_i)$ gate such that the resulting quantum state captures the effect of received wireless soft data. It works as follows: From \S\ref{s: gates}, we note that an $R_y(\theta)$ gate transforms $\ket{0}$ to $ cos \frac{\theta}{2} \ket{0} + sin \frac{\theta}{2} \ket{1}$, and so the resulting probability of a qubit $q_i$ being in state $\ket{1}$ becomes $sin^2(\frac{\theta_i}{2})$. On the other hand, we can compute the probability of a received bit being 1 using the likelihood estimation of received soft data for various modulations and channels. For instance, for BPSK-modulated data transmitted over an AWGN channel with noise variance $\sigma^2$, this probability is $1/(1~+~e^{-2y_i/\sigma^2})$. By equating these two probabilities, we obtain $\theta_i \forall i$, thus ensuring proper quantum initialization into the received soft data. Let this initialization step be the transformation:
\begin{equation}
    \ket{0}^{\otimes N}\longrightarrow U_I\ket{0}^{\otimes N} = \ket{\psi_{I}}
    \label{eqn: init}
\end{equation}
where $\ket{\psi_I}$ is our quantum state after initialization step, and $U_I = \tens{\forall i} R_y(\theta_i)$ is its corresponding unitary matrix operator.

\begin{figure*}[ht]
\centering
\begin{subfigure}[b]{0.29\linewidth}
\includegraphics[width=0.9\linewidth]{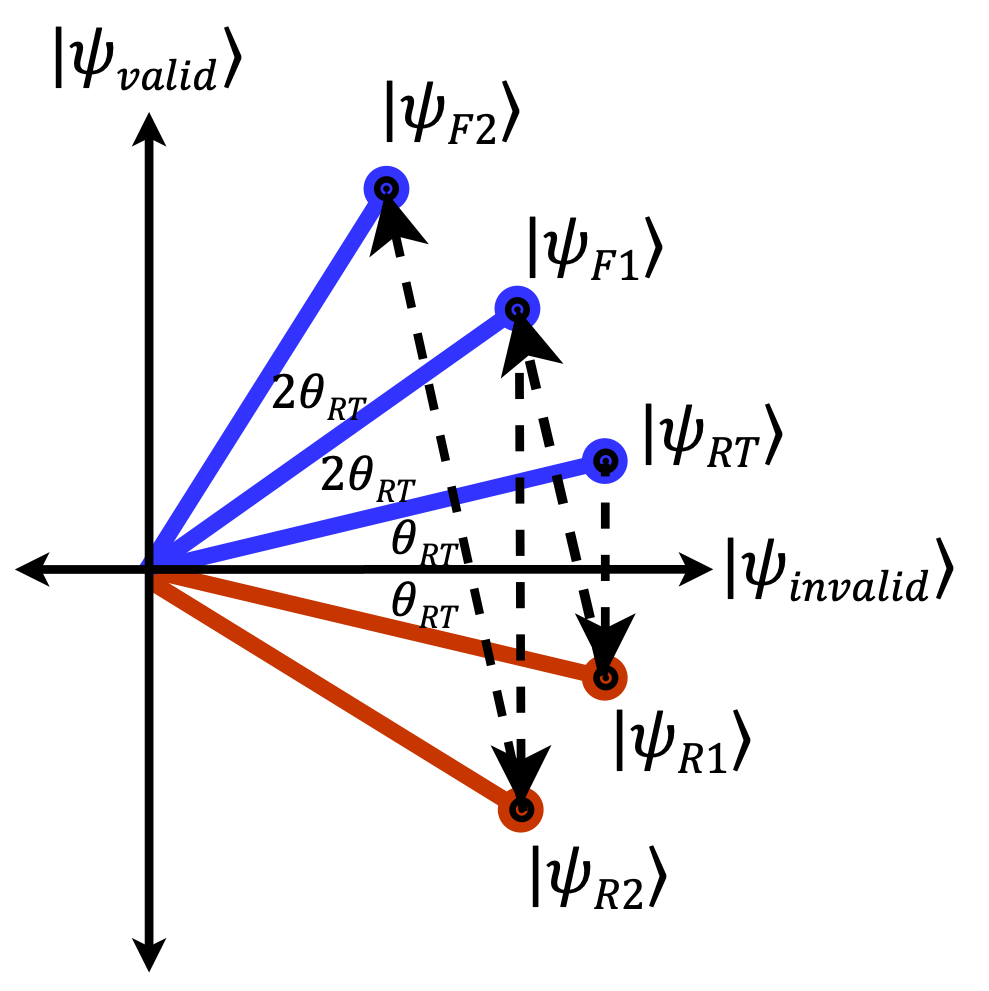}
\end{subfigure}
\hfill
\begin{subfigure}[b]{0.7\linewidth}
\includegraphics[width=\linewidth]{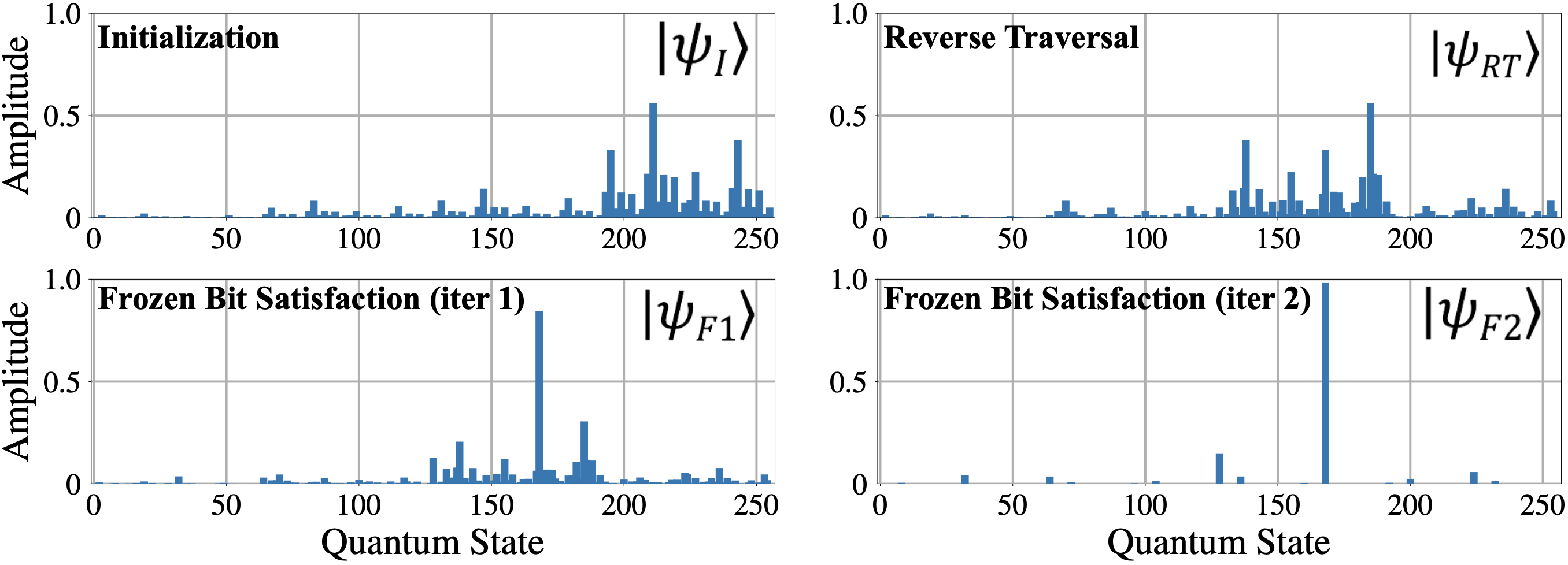}
\end{subfigure}
\caption{\textit{Left.} Frozen Bit Satisfaction via Amplitude Amplification, a technique to evolve $\ket{\psi_{RT}}$ into $\ket{\psi_{valid}}$ (see \S\ref{s: algorithm}). \textit{Right.} The working process of \systemname{} with real data at SNR 1.5~dB, showing that the ML solution (state $\ket{160}$) reaches highest decoding probability (\textit{bottom-right} plot). Quantum State indicates the configuration $\ket{q_7q_6q_5q_4q_3q_2q_1q_0}$ in integer format.}
\label{fig: ampamp}
\end{figure*}

\begin{figure*}
\centering
\includegraphics[width=0.93\linewidth]{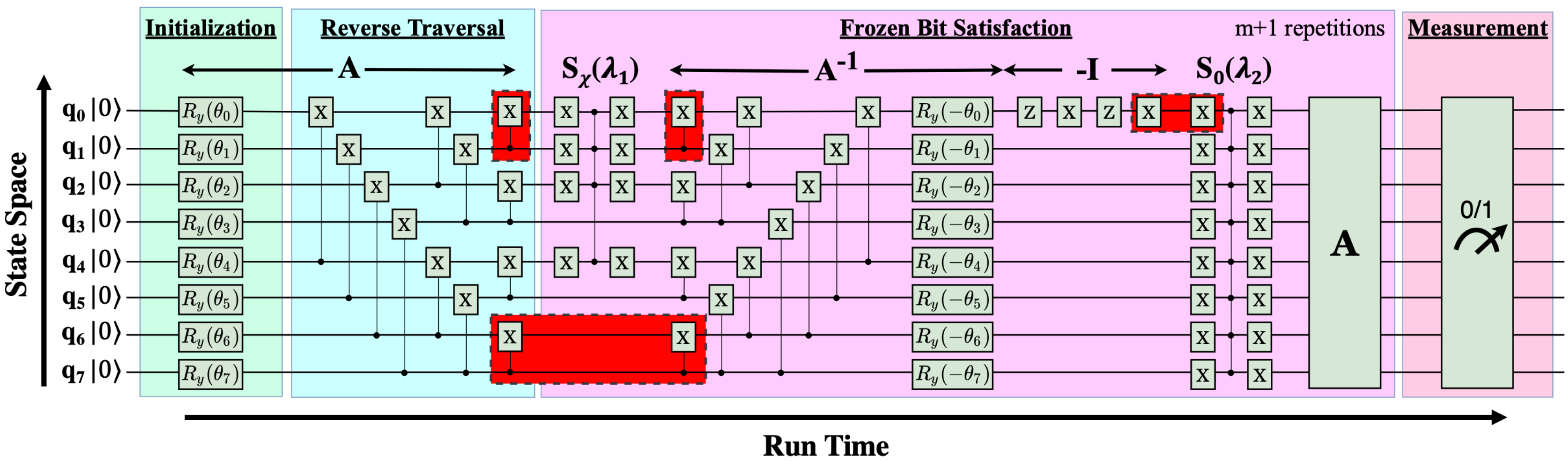}
\caption{\systemnames{} quantum circuit design for decoding the example Polar code in Fig.~\ref{fig: encoder}. $\theta_i$ is computed using the recieved soft data $y_i$ (see \S\ref{s: algorithm}). $\lambda_1$ and $\lambda_2$ are MCPhase gate angles. Gates in red/enclosed in dotted lines are removable (\S\ref{s: circuit}).}
\label{fig: circuit}
\end{figure*}

\textbf{Reverse Traversal.} Since the received soft data \textbf{y} corresponds to bits in \textbf{x}, the quantum state $\ket{\psi_I}$ resulted from the above initialization step represents bits in \textbf{x} as well. Therefore, we next evolve this quantum state from representing \textbf{x} to representing \textbf{u} (wherein user data is located), via a reverse traversal of the Polar code's Generator matrix functionality. This process can be visualized in Fig.~\ref{fig: encoder} as moving from \textbf{x} to \textbf{u} (right to left). We realize this reverse traversal evolution of quantum state by using CNOT gates, which are quantum analogies of classical XOR gates. For every XOR operation the Polar code encoder performs, we use a CNOT gate between appropriate qubits to reverse the effect of such operation. Let this reversal traversal step be the transformation:
\begin{equation}
    \ket{\psi_{I}}\longrightarrow U_{RT}\ket{\psi_{I}} = \ket{\psi_{RT}}
    \label{eqn: reversetraversal}
\end{equation}
where $\ket{\psi_{RT}}$ is the resulting quantum state after the reversal traversal step, and $U_{RT} = \otimes_{\forall 
 XOR(i, j) \in \textbf{G}_N} CNOT_{i, j}$ is its corresponding unitary matrix operator.

\textbf{Frozen Bit Satisfaction.} After the reverse traversal step, our quantum state $\ket{\psi_{RT}}$ has reached \textbf{u} wherein user data and frozen bits are located (see Fig.~\ref{fig: encoder}). However, $\ket{\psi_{RT}}$ yet does not have the knowledge of frozen bits being zero, which we must ensure to correctly decode. As such, $\ket{\psi_{RT}}$ can be thought of as a superposition of valid and invalid subspaces. We call a subspace as \textit{valid} if and only if the qubits representing frozen bits are in state $\ket{0}$, \textit{invalid} otherwise. We can express this superposition as:
\begin{equation}
    \ket{\psi_{RT}} = cos (\theta_{RT})\ket{\psi_{invalid}} + sin (\theta_{RT})\ket{\psi_{valid}}
    \label{eqn: psirt}
\end{equation}
for some $\theta_{RT}$. $\ket{\psi_{valid}}$and $\ket{\psi_{invalid}}$ are valid and invalid subspaces of $\ket{\psi_{RT}}$ respectively, and so they must be orthogonal (see Fig.~\ref{fig: ampamp} \textit{left}). Our goal is now to evolve the state $\ket{\psi_{RT}}$ such that the resulting quantum state becomes $\ket{\psi_{valid}}$, thus ensuring that our decoding agrees with the frozen bit conditions. This goal can be visualized as rotating $\ket{\psi_{RT}}$ onto $\ket{\psi_{valid}}$ in Fig.~\ref{fig: ampamp} \textit{left}. We achieve this rotation via a procedure known as amplitude amplification \cite{brassard2002quantum}. It works as follows (see Fig.~\ref{fig: ampamp} \textit{left}): Set~$\ket{\psi} = \ket{\psi_{RT}}$. Reflect $\ket{\psi}$ about $\ket{\psi_{invalid}}$, evolving into a quantum state $\ket{\psi_{R1}}$, then reflect $\ket{\psi_{R1}}$ about $\ket{\psi_{RT}}$, evolving into a quantum state $\ket{\psi_{F1}}$. These pair of reflections constitute an iteration. Now set $\ket{\psi} = \ket{\psi_{F1}}$, and repeat the procedure iteratively, so that eventually our quantum state ends up near $\ket{\psi_{valid}}$. Each iteration achieves a $2\theta_{RT}$ rotation (see Fig.~\ref{fig: ampamp} \textit{left}), and so by performing $m = \lceil\frac{\pi}{4\theta_{RT}} - 0.5\rceil$ number of iterations alongside a slow final rotation (\textit{i.e.,} one $< 2\theta_{RT}$ rotation), $\ket{\psi_{RT}}$ evolves into $\ket{\psi_{valid}}$ with high certainty \cite{brassard2002quantum}. Let this frozen bit satisfaction step be the transformation:
\begin{equation}
    \ket{\psi_{RT}}\longrightarrow U_{FBS}\ket{\psi_{RT}} = \ket{\psi_{valid}}
\end{equation}
$U_{FBS} = \{-AS_0(\lambda_2)A^{-1}S_\chi(\lambda_1)\}^{m+1}$ is this step's unitary operator where $A = U_{RT}U_I$ (see Eqs.~\ref{eqn: init}, \ref{eqn: reversetraversal}). $S_\chi$ and $-AS_0A^{-1}$ are operators that perform reflections about $\ket{\psi_{invalid}}$ and $\ket{\psi_{RT}}$ respectively. In particular, $S_\chi(\lambda_1)$ and $S_0(\lambda_2)$ work by transforming $\ket{\psi_{valid}}$ to $e^{i\lambda_1} \ket{\psi_{valid}}$ and $\ket{0}^{\otimes N}$ to $e^{i\lambda_2}\ket{0}^{\otimes N}$ respectively. For the first \textit{m} iterations, $\lambda_1 = \lambda_2 = \pi$, and for the last slow rotation $\lambda_1 = cos^{-1}(-cot ( 2\theta_{RT})/tan ((2m+1)\theta_{RT}))$ and $\lambda_2 = 2tan^{-1}(-cot(\lambda_1)/cos(2\theta_{RT}))$. We omit the derivation of these values in this paper, see \cite{brassard2002quantum} and  \cite{long2002phase} for details.

After the frozen bit satisfaction step, our quantum state has now reached $\ket{\psi_{valid}}$. Measuring the qubits in this $\ket{\psi_{valid}}$ state returns the Maximum-Likelihood (ML) solution with highest probability. We now demonstrate this with real data, in Fig.~\ref{fig: ampamp} \textit{right}. The Initialization plot shows the distribution of amplitudes in $\ket{\psi_{I}}$, which is essentially the distribution of received soft data. As $\ket{\psi_{I}}$ evolves into $\ket{\psi_{RT}}$, we note in the Reverse Traversal plot that these amplitudes are \textit{shifted} among the states without changing magnitude, as is how XOR/CNOT operations work (\textit{cf.} Eqs.~\ref{eqn: cnot_0}, \ref{eqn: cnot}). The \textit{Frozen Bit Satisfaction} plots show how the ML solution (state $\ket{160}$) is amplified from iteration to iteration, as we evolve $\ket{\psi_{RT}}$ into $\ket{\psi_{valid}}$, reaching a near one decoding probability.

\begin{figure*}
\centering
\includegraphics[width=\linewidth]{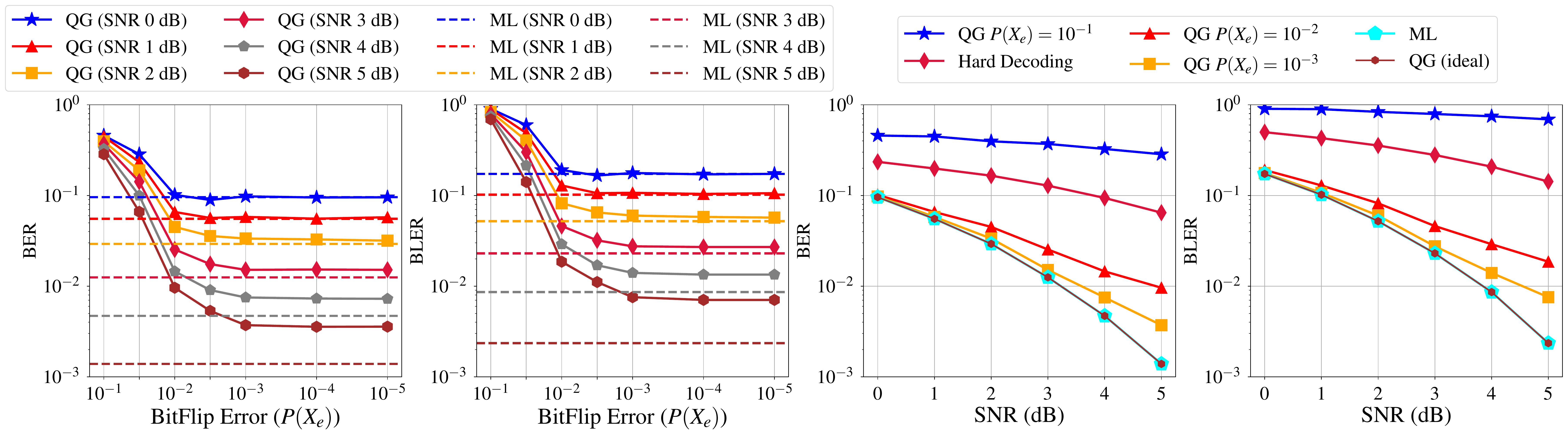}
\caption{\systemnames{} decoding performance in ideal and noisy quantum gate simulations.}
\label{fig: result}
\end{figure*}

\subsection{\systemname{}: Quantum Circuit}
\label{s: circuit}

In Fig.~\ref{fig: circuit}, we illustrate \systemnames{} quantum circuit design, which is a space-time diagram showing qubits (horizontal wires) and quantum gates that act upon them. The Initialization and Reverse Traversal blocks show $R_y(\theta)$ gates and CNOT gates (dot is control qubit) respectively. The placement of these CNOT gates is the reverse sequence of XOR gates the encoder performs (see Fig.~\ref{fig: encoder}), thus mimicking the Polar code's Generator matrix functionality. We next construct the frozen bit satisfaction operator $U_{FBS} = \{-AS_0(\lambda_2)A^{-1}S_\chi(\lambda_1)\}^{m+1}$ as follows: $S_\chi (\lambda_1)$ operator is realized via an MCPhase($\lambda_1$) gate sandwiched between X gates as shown in Fig.~\ref{fig: circuit}, and it operates only on qubits representing frozen bits. This ensures that $\ket{\psi_{valid}}$ (states with frozen bits as zero) is transformed to $e^{i\lambda_1}\ket{\psi_{valid}}$ (\textit{cf.} \S\ref{s: gates}, \S\ref{s: algorithm}). $A^{-1}$ operator is reverse of $A$ with negative $R_y$ angles. This is because the inverse of $R_y (\theta)$ is $R_y(-\theta)$ and CNOT gates are self-inverse gates. The negative sign in $U_{FBS}$ is realized via the standard ZXZX gate sequence, which can be applied onto any qubit. Similar to $S_\chi (\lambda_1)$, $S_0 (\lambda_2)$ is realized via MCPhase($\lambda_2$) gate sandwiched between X gates, and it operates on all qubits, therefore ensuring that $\ket{0}^{\otimes N}$ is transformed into $e^{i\lambda_2}\ket{0}^{\otimes N}$ (\textit{cf.} \S\ref{s: gates}, \S\ref{s: algorithm}). The frozen bit satisfaction block is repeated for $m+1$ iterations. We also optimize the circuit by removing consecutive CNOT gates and consecutive X gates, which cancel the effect of each other. CNOT gates surrounding two frozen bits are also removed as zero-inputs result in zero-outputs.

\section{Preliminary Results and Discussion}
\label{s: eval}

We have evaluated \systemname{} for eight bit, 0.5 rate Polar codes in both ideal and noisy quantum gate simulations. Our noisy simulations consider \textit{BitFlip} quantum noise, which applies an X-gate operation before each gate in Fig.~\ref{fig: circuit} with a pre-defined probability of $P(X_e)$. We limit the number of iterations in frozen bit satisfaction step to a maximum of five, and round up $R_y$ gate angles to nearest integer factor of $\pi/128$. Our evaluation dataset consists $5\times 10^6$ problems, where each problem is solved $10^3$ times, and the solution with most occurrences is considered for evaluation.

We see in Fig.~\ref{fig: result} \textit{left} and \textit{mid-left} plots the bit error rate (BER) and block error rate (BLER) performance of \systemname{} as the quantum noise varies. At regions of high noise (\textit{i.e.,} $P(X_e) > 10^{-2}$), the BER/BLER performance of \systemname{} is far away from the ML performance at all wireless channel SNRs, whereas when $P(X_e)$ is below $10^{-2}$, the performance of \systemname{} improves and then saturates. We also note that the performance gap between \systemname{} and the ML decoder increases with increase in SNR. This implies that when $P(X_e)$ is below $10^{-3}$, rather than quantum gate noise the aforementioned algorithmic approximations limit performance. We next see in Fig.~\ref{fig: result} \textit{mid-right} and \textit{right} plots the performance of \systemname{} as the wireless channel SNR varies. The figure shows that \systemname{} achieves the ML performance in the ideal scenario (no noise, no iteration limit, no $R_y$ angles quantization), whereas at $P(X_e)$ of $10^{-3}$ \systemname{} achieves about half an order of higher BER/BLER magnitude than the ML decoder at SNR 5~dB.

Our simulation results are encouraging, demonstrating the promising performance of quantum gate computation. Scaling \systemname{} to long 1,024 bits 5G-NR Polar codes via a hybrid classical--quantum approach and its implementation on real quantum computers is potential future work direction. Issues surrounding today's quantum technology such as limited qubit count and noise may challenge direct implementations of \systemname{}, which an eventual full system design must overcome. In particular, we must tailor \systemname{} to hardware at hand via a full stack software-hardware co-design. The ideas we propose here will also benefit from significant research being dedicated towards fault-tolerant quantum computing, which in turn may enable applicability of Polar codes in NextG cellular traffic channels.

\section*{Acknowledgements}

This research is supported by National Science Foundation (NSF) Award CNS-1824357. We gratefully acknowledge a gift from InterDigital Corporation.

\bibliographystyle{abbrv}
\bibliography{references}

\end{document}